\documentclass[12pt]{article}

\usepackage[margin=1in]{geometry} 
\usepackage{amsmath} 
\usepackage{graphicx}
\graphicspath{ {Pictures/} }

\begin{document}
\title{Comparison of DAMA/LIBRA and DM Ice Results using Information Theory to Rule out Dark Matter Claims}
\author{D. Cline \& M. Simpson \\ UCLA Astroparticle Physics Division}
\date{\today}
\maketitle
\begin{abstract}
  We study the details of the DAMA/LIBRA results and compare those with the recent published DM Ice results of ICE Cube. In various recent papers, it was shown that the $^{40}$K peak on DAMA/LIBRA data leaves no room for a  Dark Matter signal in the bulk of the data. Using Information Theory for the different types of detection environments, we show that annual variation calculations and the DM Ice data reinforce the claims that the DAMA/LIBRA detector is not observing Dark Matter WIMPs.
\end{abstract}

\section{Introduction}
\indent In this paper, we study the DAMA/LIBRA data [1] and compare it to background calculations done by other groups  and to recent DM ICE [2] data. The major question we seek to answer is if there is an absence of any WIMP-like signal in the bulk of the data. Recently, several works have shown that the annual variation in the DAMA data can be produced by neutron and muon interactions. Nearly exact fits have been produced in tests done by J. Davis [3] (Fig. 1) and can be compared to the earlier fits that use signal measured on the ICARUS detector at the LGNS [4] (Fig. 2). In a Dark Matter model, the annual variation of the signal is due to the different WIMP interactions as the Earth goes around the sun. Therefore, there must be considerable WIMP interactions by the bulk of the data. See Fig. 3.

\section{Annual Variation of WIMPs [9]}

\begin{center}
\includegraphics[scale=0.9]{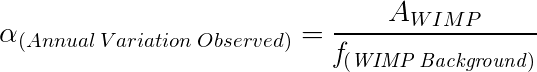}
\end{center}

\indent The real annual variation of WIMPs tends to be misunderstood by several groups in this field. Annual variation comes from the formula shown on the previous page. Most people (including DAMA) only report on the value of $\alpha$. However, the fraction of WIMPs (or other words dark matter) in the data plays a crucial role in finding the actual annual variation. In work done by J. Davis, he shows that this annual variation observed from DAMA can easily be matched with their simulations of Dark Matter as well as Neutrinos+Muons that would be observed in the same case. This justifies that there must be more to wimp annual variation than what is observed. There must be a $f_{(WIMP\: Background)}$ component that gives the actual $A_{WIMP}$. Figure 1 shows the comparison of Davis' model and the DAMA data [3]. In their model they have shown how contributions from Neutrinos and Muons lead to the creation of neutrons in a detector and can create a fit similar to what DAMA sees in their annual modulations. If we are to find an annual variation of WIMPs we must use more than the annual variation we observe in these detectors. We must account for the fraction of WIMPs that are in our background on top of other background contributions like solar effects and detector impurities. In essence the effect due to motion around the sun is very small and undetectable unless one has a system with many Dark Matter particle interactions. We find that this is not the case in DAMA.  \\

\begin{figure}[ht]
\centering
\includegraphics[width=\textwidth]{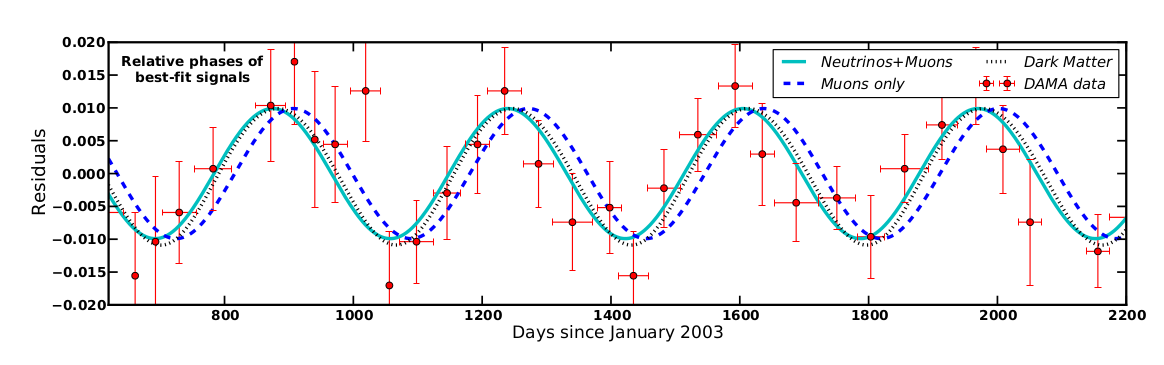}
\caption{\label{first}Comparison of annual variation models for the DAMA data. The cyan line is the model proposed by J. Davis [3] that is composed of neutrons produced by solar neutrinos and atmospheric muons. Both the Dark Matter and J. Davis model fit to the DAMA modulation equally well.}
\end{figure}

\section{The Fraction of Dark Matter Interactions in the Detector}
In order for Dark Matter interactions to cause a slight annual variation from the small effect of the earth going around the sun, there should be a measurable level of these interactions. We call this measurable level $f$ for fraction of dark matter and is the denominator of our annual variation equation. Work done by Katherine Freese et al shows the importance of the Dark Matter Halo's structure in determining the annual variation one would observe in a detector [9]. 

\begin{center}
\includegraphics[scale=0.5]{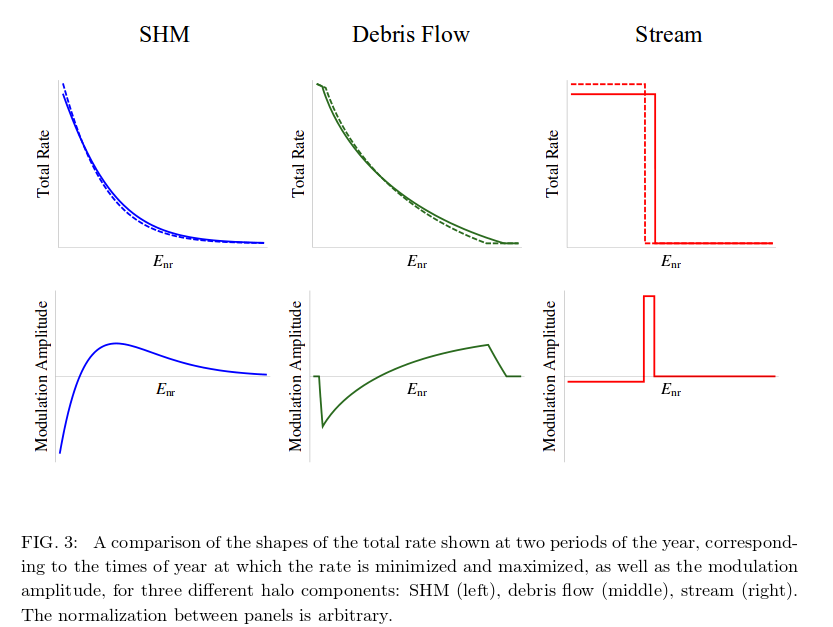}
\end{center}

In order to see a significant annual variation from dark matter alone, we would need to be observing a very dense background of WIMPs or have an extremely sensitive detector. By the calculations of Freese et al, these sensitivities would be beyond the bounds of most detectors including DAMA. Freese et al make a point in their paper to to simulate the various conditions required to observe annual modulation due to dark matter. The following figure is an image they used to show the various cases in which annual modulation was maximized and minimized. In their paper they concluded that you would need sizable recoil energies to witness the variation in amplitudes. Looking back to Pradler's work, it would seem plausible for neutrons created within the detector to cause these large recoils [5]. \\

\indent Figure 2 shows the results of Ralston whose group analyzed the DAMA data to search for these WIMP interactions [4]. The time dependent neutron fit they study makes a good fit to the DAMA data.
In fact, it seems that DAMA has also made the same calculation (red curve) and is aware of this contribution to variation they see. To add to this, the new DM ICE preliminary data seems consistent with these dominant neutron rates being observed. All results indicate that $f$ is very small and that DAMA is not witnessing annual variation dominated by dark matter.

\begin{figure}[ht]
\centering
\includegraphics[width=\textwidth]{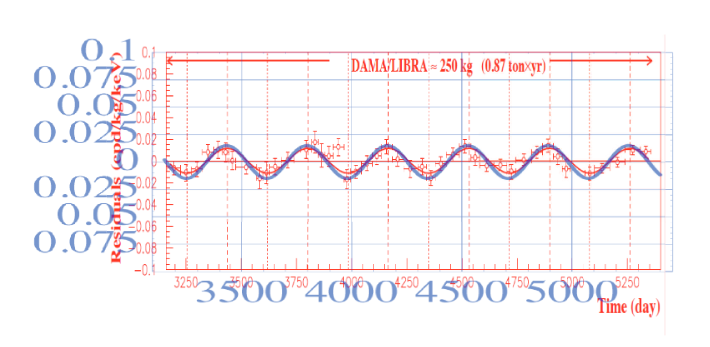}
\caption{\label{second}Overlay of time dependence depicted by Gran Sasso underground neutron rate (blue line) with the time dependence of DAMA/LIBRA signal region (red) [4]}
\end{figure}

\indent In order to calculate $f$ it is necessary  to know the level of radioactive background in DAMA. The most important component to know is the contribution from $^{40}K$. The radioactive decay comes at 3 keV, the same region that DAMA claims a strong annual variation. This is just a coincidence. In the past, the addition of the $^{40}K$ signal to the other backgrounds has been done imprecisely. However, the John Hopkins group, which uses exactly the same abundance of $^{40}K$ produced by DAMA, has done excellent work in finding its contribution to background [5]. The results are shown in Fig. 3 and indicate a small value of $f$ as well. The DAMA group has done this same calculation also (red curve) and their peak is only (different) off the John Hopkins' data by a small amount. DAMA has replied to these claims done by the John Hopkins group saying that DAMA's contamination by $^{40}K$ is actually 13 ppb and calculations done by this group were done at 20 ppb and are therefore wrong. This is not the case since Pradler stated that his group tested the material for the entire range of 1-100 ppb and found that at 13 ppb DAMA would need a signal modulation fraction of over 20\%. This is additional evidence towards the DAMA annual variation and signal being due to background instead of dark matter [Fig. 3]. \\
\indent So to recap, table 1 shows that a considerable background simulation has been carried out by several groups including DAMA that agrees with the data without any WIMP interactions, f is close to zero making a Wimp very large as shown in Fig 4.

\begin{table}[htbp]
\centering
\caption{Current Estimates of the Background Level in DAMA}
  \begin{tabular}{ | l | c | r |}
    \hline
    V. Kudryavtsev     & Detailed calculation of background & Ref 11 \\
    et al          & agrees with DAMA data              &       \\ \hline
    Peter          & Made similar background calculation& Unpublished \\
    Smith          &  with DAMA NaI                     &       \\ \hline 
    Praedler       & Correctly put in $^{40}$K into     & Ref 5 \\
    et al          & background \& made calculation past&       \\
                   & initial peak                       &       \\ \hline
    DAMA Group          & Made a calculation of $^{40}$K background   & Ref 7 \\ 
                   & long ago (unpublished)             &       \\
    \hline
  \end{tabular}
\end{table}

\begin{figure}[ht]
\centering
\includegraphics[width=\textwidth]{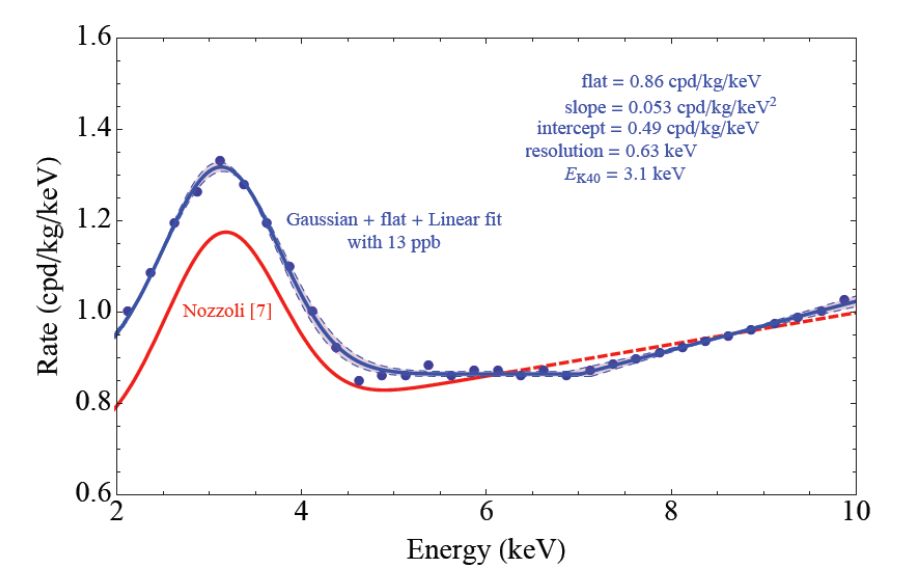}
\caption{The rate of interactions in DAMA. The red line is an actual DAMA result found by John Hopkins group. The blue line is a simulation - the dots are DAMA data (See text and Table 1.}
\end{figure}

\section{Interpreting Signal from all Sources of Background via Information Theory}
\indent There are several contributions to background that have been overlooked in the DAMA data. For one, Davis had shown that the bold statement s made for WIMPs in the detector could  have been made from by neutrinos creating neutrons. In other cases, Freese [9] had shown that the type of DM halo and WIMP fraction were in direct correlation with annual variation and therefore affected how many WIMPs you should expect to detect. Furthermore, many assumptions tend to be made in detectors such as DAMA in which signals you can keep versus which ones are ignored. \\
\indent To avoid mistakes, misrepresentations of data, and bold claims on a WIMP signal, Information Theory provides a method that looks at all sources of noise-to-signal information [Fig. 3]. It quantizes it into a these types of mathematical bins that allows computers do better differentiate signals from noise. It is a method that goes over looked in most direct detection experiments and provides better decisions for whether this data is "signal" or "noise". Cousins gives situations in which using Information Theory instead of typical $\chi^{2}$ methods gives sharper/more-defined signals through a set of background elements. Though it isn't perfect in every case (he differentiates the scenarios where $\chi^{2}$ can have better results), Information theory is a method that goes overlooked in these detector experiments and is key to comparing one detectors set of data do a different detector's data. One such detector where DAMA should be compared is the new data given by DM-Ice. \\
\indent Numerous sources of information are ignored in the DAMA experiment such as neutron generation within detectors, solar effects, the correct formula for annual variation due to a WIMP background, an understanding of the $^{40}K$ contribution, etc. The power in information theory lies in that it accounts for all these sources of information and puts it in a quantifiable form that can be directly applied to distinguishing signal from background. It can directly compare one experiments data, for example DAMA, with the next, like DM-Ice, in a universal way.

\section{Annual Variation due to DM and Confirmed Lack of Signal in DAMA from New DM-Ice Results}
\indent As validation for the unlikelihood of DAMA having detected a Dark Matter signal beyond any of these background components, DM-Ice has released new data from as of August 2014. Their detector, regarded as a successor to this previous generation of direct detectors, has taken data showing that if they found a WIMP signal it would still be shrouded by their detector's background. Figure 5 is taken from DM-Ice's first data. Expected DM signals would give a peak in the 2-4 keV range and would need to be above the total background (red line). However, the observed data is actually below this background altogether. Considering that DM-Ice as more sensitive as DAMA, these results reaffirm a lack of dark matter signal in the DAMA data. Therefore, WIMP signals in the DAMA data must have come from other unaccounted sources such as underestimation in $^{40}K$ background, creation of neutrons within the detector by solar influences, and Neutrino+Muon backgrounds.

\begin{figure}[ht]
\centering
\includegraphics[width=\textwidth]{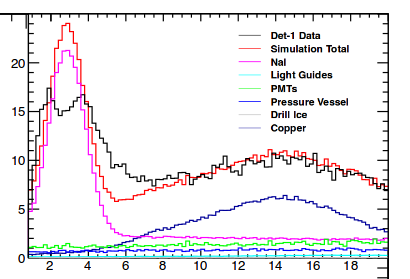}
\caption{DM-Ice first run data. The keV range (x-axis) we expect to find a WIMP signal is in the low sharp peaked region. The counts/day/keV/kg show the data being shrouded by the detector's background contributions.}
\end{figure}

\newpage
\section{Conclusions}
In general, a more careful understanding of background contribution and its application to the correct annual variation of WIMPs formula should be required from DAMA before any of there results are taken seriously. If these points are readdressed within DAMA's data and appear conclusive with future experiments such as future DM-Ice results then these claims may lead to new results. Until then, this data should not be trusted.

\newpage
\section{References}

$[1]$ R Bernabei et al., Eur.Phys.J. C73, 2648 (2013) 1308.5109. \\ 
$[2]$ DM-Ice Collaboration: J. Cherwinka et al., arXiv:1401.4804. 13 Aug, 2014. \\ 
$[3]$ J. Davis arXiv:1407.1052v2. Aug 2014 \\ 
$[4]$ J. Ralston (2010) arXiv:1006.5255 28 Jun 2010 \\ 
$[5]$ J. Pradler et al (2013) John hopkins arXiv:1210.5501 \\
$[6]$ R. Bernabei et al (2012) DAMA arXiv:1210.6199v1 \\ 
$[7]$ J. Pradler arXiv:1210.7548v1 \\
$[8]$ D. Cline (2013) arXiv:1308.3477 \\
$[9]$ K. Freese et al (2013) arXiv:1209.3339v3 \\
$[10]$ R. D. Cousins (1994)\\
$[11]$ V. Kudryavtsev et al, Astropart.Phys. 33, 91 (2010)\\

\end{document}